# New orbits of irregular satellites designed for the predictions of stellar occultations up to 2020, based on thousands of new observations

A. R. Gomes-Júnior[1]*, M. Assafin[1,†], L. Beauvalet[2,3], J. Desmars[4],
R. Vieira-Martins[1,2,5,†], J. I. B. Camargo[2,5], B. E. Morgado[1,2] F. Braga-Ribas[2,6]

[1] Observatório do Valongo/UFRJ, Ladeira Pedro Antônio 43, CEP 20.080-090 Rio de Janeiro - RJ, Brazil
[2] Observatório Nacional/MCTI, R. General José Cristino 77, CEP 20921-400 Rio de Janeiro - RJ, Brazil
[3] Observatoire de Paris/SYRTE, 77 Avenue Denfert Rochereau 75014 Paris, France
[4] IMCCE, Observatoire de Paris, PSL Research University, CNRS, Sorbonne Universités, UPMC Univ Paris 06, Univ. Lille, 77 Av. Denfert-Rochereau, 75014 Paris, France
[5] Laboratório Interinstitucional de e-Astronomia - LIneA, Rua Gal. José Cristino 77, Rio de Janeiro, RJ 20921-400, Brazil
[6] Federal University of Technology - Paraná (UTFPR / DAFIS), Rua Sete de Setembro, 3165, CEP 80230-901, Curitiba, PR, Brazil
[†] Affiliated researcher at Observatoire de Paris/IMCCE, 77 Avenue Denfert Rochereau 75014 Paris, France

**ABSTRACT**

Gomes-Júnior et al. (2015) published 3613 positions for the 8 largest irregular satellites of Jupiter and 1787 positions for the largest irregular satellite of Saturn, Phoebe. These observations were made between 1995 and 2014 and have an estimated error of about 60 to 80 mas. Based on this set of positions, we derived new orbits for the eight largest irregular satellites of Jupiter: Himalia, Elara, Pasiphae, Carme, Lysithea, Sinope, Ananke and Leda. For Phoebe we updated the ephemeris from Desmars et al. (2013) using 75% more positions than the previous one. Due to their orbital characteristics, it is common belief that the irregular satellites were captured by the giant planets in the early Solar System, but there is no consensus for a single model explaining where they were formed. Size, shape, albedo and composition would help to trace back their true origin, but these physical parameters are yet poorly known for irregular satellites. The observation of stellar occultations would allow for the determination of such parameters. Indeed Jupiter will cross the galactic plane in 2019-2020 and Saturn in 2018, improving a lot the chances of observing such events in the near future. Using the derived ephemerides and the UCAC4 catalogue we managed to identify 5442 candidate stellar occultations between January 2016 and December 2020 for the 9 satellites studied here. We discussed how the successful observation of a stellar occultation by these objects is possible and present some potential occultations.

## 1 INTRODUCTION

Irregular satellites revolve around giant planets at large distances in eccentric, highly inclined and frequently retrograde orbits. Because of these peculiar orbits, it is largely accepted that these objects did not form by accretion around their planet, but were captured in the early Solar System (Sheppard 2005).

There is no consensus for a single model explaining where the irregular satellites were formed. Ćuk & Burns (2004) showed that the progenitor of the Himalia group may have originated in heliocentric orbits similar to the Hilda asteroid group. Sheppard

(2005) stated that the irregular satellites may be some of the objects that were formed within the giant planets region.

Grav et al. (2003) and Grav & Bauer (2007) showed that the irregular satellites from the giant planets have their colors and spectral slopes similar to C-, D- and P-type asteroids, Centaurs and trans-neptunian objects (TNOs). This suggests that they may have come from different locations in the early Solar System.

Sheppard (2005) and Jewitt & Haghighipour (2007) also explored the possibility that the irregular satellites originated as comets or TNOs. TNOs are highly interesting objects that, due to their large heliocentric distances, may be highly preserved with physical properties similar to those they had when they were formed (Barucci et al. 2008). This is even more true for the smaller objects, since in principle larger sizes favour physical differenti-

* E-mail: altair08@astro.ufrj.br

**Table 1.** Estimated diameter of the satellites and correspondent apparent diameter

| Satellite | Diameter of the satellites | | Ref. |
| | mas[a] | km | |
|---|---|---|---|
| Ananke | 8 | 29 | 1 |
| Carme | 13 | 46 | 1 |
| Elara | 24 | 86 | 1 |
| Himalia | 41 | $(150 \times 120) \pm 20$[b] | 2 |
| Leda | 5 | 20 | 1 |
| Lysithea | 10 | 36 | 1 |
| Pasiphae | 17 | 62 | 1 |
| Sinope | 10 | 37 | 1 |
| Phoebe | 32 | $212 \pm 1.4$[b] | 3 |

References: (1) Rettig et al. (2001); (2) Porco et al. (2003); (3) Thomas (2010).

[a] Using a mean distance from Jupiter of 5 AU, from Saturn of 9 AU and from Neptune of 30 AU.

[b] From Cassini observations.

ation processes in the body and vice-versa. However, due to the distance, the smaller TNOs from this region are more difficult to observe. Thus, if irregular satellites - or at least a few of them - do share a common origin with small TNOs, and since these objects are situated at much closer heliocentric distances now, this gives a unique chance of observing and studying representatives of this specific TNO population in much greater detail than could ever be possible by direct observation of this population in the Kuiper Belt.

Phoebe is the most studied irregular satellite. Clark et al. (2005) suggest that its surface is probably covered by material of cometary origin. It was also stated by Johnson & Lunine (2005) that if the porosity of Phoebe is 15%, Phoebe would have an uncompressed density similar to those of Pluto and Triton.

Gomes-Júnior et al. (2015) (which shall be referred as GJ15 hereafter) obtained 6523 suitable positions for 18 irregular satellites between 1992 and 2014 with an estimated error in the positions of about 60 to 80 mas. For some satellites the number of positions obtained is comparable to the number used in the numerical integration of orbits by the JPL (Jacobson et al. 2012). They pointed out that the ephemeris of the irregular satellites have systematic errors that may reach 200 mas for some satellites.

We present in this paper new numerical integration of the orbits of the 8 major irregular satellites of Jupiter (Himalia, Elara, Pasiphae, Lysithea, Carme, Ananke, Sinope and Leda) using only the positions obtained by GJ15 (see Sec. 2). For Phoebe, we updated the ephemeris of Desmars et al. (2013) with the observations of GJ15, Peng et al. (2015), observations from Minor Planet Center and observations from Flagstaff.

Phoebe, being the most studied object with a good measured size, can be used to calibrate and evaluate the technique for similar objects. Up to date, no observation of a stellar occultation by an irregular satellite was published. Since their estimated sizes are very small (see Table 1), this may have discouraged earlier tries. But, in fact, given their relatively closer distances as compared to TNOs and Centaurs, and considering the precision of our ephemeris and of star positions, we can now reliably predict the exact location and instant where the shadow of the occultation will cross the Earth.

In section 2, we present the new determination of the orbits. In section 3, we present the predictions of the stellar occultations by irregular satellites, including some tests made to check the accuracy of the predictions. The final discussion is presented in section 4.

## 2 ORBIT COMPUTATIONS

GJ15 published 3613 precise positions for the 8 largest irregular satellites of Jupiter from observations made at the Observatório do Pico dos Dias (OPD), Observatoire Haute-Provence (OHP) and European Southern Observatory (ESO) between 1995 and 2014.

Here we compute new orbits based on the observations published in GJ15. First, because the reduction was made with a consistent and precise stellar catalogue and with a robust astrometry (PRAIA, Assafin et al. 2011). Second, besides recent observations, this consistent set of numerous and precise positions covers many orbital periods at many distinct orbital plane sights, allowing to fully constrain the orbit for the short time span explored in this work. For these reasons, only this set of positions was used for the satellites of Jupiter.

Due to the context of this work regarding to stellar occultations, the orbit fitting procedures used aimed primarily to derive precise ephemerides for the near future. Technically, the procedures easily allow for the continuous addition of more observations (old, new) aiming at refining the orbit fits.

### 2.1 Special-Tailored Ephemerides (STE) for Jupiter irregular satellites

The last observations used to develop current JPL ephemeris of the irregular satellites of Jupiter were obtained in 2012 (Jacobson et al. 2012). As a result, the errors in the JPL ephemeris for the current epoch may be probably too large to prevent accurate predictions of stellar occultations without any corrections.

Our numerical model describes the dynamical evolution of the irregular satellites of Jupiter in a jovicentric reference frame. The satellites are submitted to the influence of the Sun and the main bodies of the solar system (from Mercury to Pluto, plus the Moon), as well as those of the Galilean satellites and the first harmonics of Jupiter's gravity field. The axes of the reference frame are expected to be those of the ICRS.

We use the following notations:

- in one dynamical family consisting of $\mathcal{N}$ irregular satellites, $i$ will stand for the one whose equation of motion we are considering, $l$ will stand for another irregular satellite in gravitational interaction belonging to the same family
- $\bullet$ $J$ Jupiter
- $\bullet$ $j$ another body of the Solar System, among the Galilean satellites, the Sun, the planets, Pluto and the Moon (14 bodies)
- $\bullet$ $M_j$ the mass of the $j$-th body, not an irregular satellite
- $\bullet$ $m_i$ the mass of the irregular satellite $i$
- $\bullet$ $\vec{r_i}$ the position of the $i$-th body with respect to the center of Jupiter
- $\bullet$ $r_{ij}$ the distance between bodies $i$ and $j$
- $\bullet$ $R_J$ the radius of Jupiter
- $\bullet$ $J_n$ the dynamic polar oblateness of the nth order for Jupiter's gravity field
- $\bullet$ $U_{\bar{l}j}$ potential generated by the oblateness of Jupiter on the satellite $l$
- $\bullet$ $\Phi_i$ is the latitude of the $i$-th satellite with respect to Jupiter's equator.

For an irregular satellite $i$, under the gravitational influence of Jupiter, the seven other irregular satellites, the regular Jovian satellites and the main bodies of the Solar System, the equation of

motion is:

$$
\begin{aligned}
\ddot{\vec{r_i}} = \ & -GM_J \frac{\vec{r_J} - \vec{r_i}}{r_{iJ}^3} - \sum_{l=1,l\neq i}^{\mathcal{N}} Gm_l \frac{\vec{r_l} - \vec{r_i}}{r_{il}^3} \\
& -\sum_{j=1}^{14} GM_j \left( \frac{\vec{r_j} - \vec{r_i}}{r_{ij}^3} - \frac{\vec{r_j} - \vec{r_J}}{r_{Jj}^3} \right) \\
& +GM_J \nabla U_{\bar{i}J} - \sum_{l=1}^{\mathcal{N}} Gm_l \nabla U_{\bar{i}j}
\end{aligned}
\tag{1}
$$

where the last term in brackets and the last term in Eq. 1 represent undirect perturbations. The oblateness potential seen by a satellite $i$ because of Jupiter is (with a similar expression for the oblateness seen by a satellite $l$):

$$
\begin{aligned}
U_{\bar{i}J} = \ & -\frac{R_J^2 J_2}{r_{iJ}^3} \left( \frac{3}{2}\sin^2\Phi_i - \frac{1}{2} \right) \\
& -\frac{R_J^4 J_4}{r_{iJ}^5} \left( \frac{35}{8}\sin^4\Phi_i - \frac{15}{4}\sin^2\Phi_i + \frac{3}{8} \right) \\
& -\frac{R_J^6 J_6}{r_{iJ}^7} \left( \frac{231}{16}\sin^6\Phi_i - \frac{315}{16}\sin^4\Phi_i + \frac{105}{16}\sin^2\Phi_i - \frac{5}{16} \right)
\end{aligned}
\tag{2}
$$

The expressions of $\nabla U$ have been developed in Lainey et al. (2004). The equations of motion are integrated with the 15th order numerical integrator RADAU (Everhart 1985) using a constant step of one day. The positions of the objects of the solar system are provided by the DE423 ephemeris (Folkner 2010), while those of the galilean satellites are provided by NOE2010 (Lainey et al. 2004). Our model was fitted to the observations through a least-squares procedure. The satellites were integrated one dynamical family at a time, to gain computing time, while losing minimum precision. Indeed, the interactions between satellites not belonging to the same dynamical family are negligible considering the short timespan of our integration.

The obtained ephemeris is hereafter referred to as STE, for Special-Tailored Ephemeris. The initial osculating elements at the origin of integration, the number and time span of the observations of each satellite are presented in Table 2.

Some methods to derive the errors of the ephemeris of irregular satellites can be found in Emelyanov (2010). In the Natural Satellites Ephemeride Server MULTI-SAT (Emelyanov & Arlot 2008) the precisions can be obtained for the satellites at any given time from the Emelyanov (2005) ephemeris updated to February 19, 2012. However, since the practical realization of the STE ephemeris is for help improving the prediction of stellar occultations by the irregular satellites in the immediate future, it is interesting to compare the STE with the other relevant ephemerides for the next few years.

We compared the STE ephemeris to the JPL for all the Jupiter satellites we fitted, until 2021. For instance, the maximum difference between 2015 and 2021 is at most 98 mas in $\Delta\alpha\cos\delta$ and 58 mas in $\Delta\delta$ for Himalia and 181 mas in $\Delta\alpha\cos\delta$ and 152 mas in $\Delta\delta$ for Carme.

Fig. 1 displays the offsets of the positions published by GJ15 for the satellite Carme in declination relative to the STE ephemeris, to Jacobson et al. (2012) jup300 JPL ephemeris and Emelyanov (2005)[1] ephemeris. The DE431 planetary ephemeris (Folkner et al.

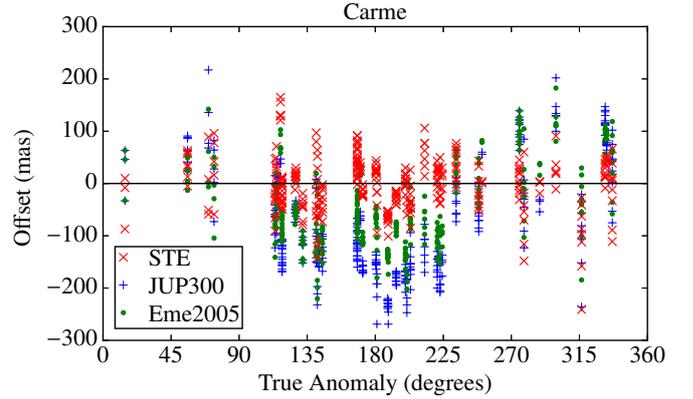

**Figure 1.** Offsets in declination of the positions published by GJ15 for Carme. The red "x" relate to the special-tailored ephemeris (rms = 51), the blue "+" to the jup300 JPL ephemeris (rms = 130) and the green dot to Emelyanov (2005) (rms = 92). As expected, the ephemeris offsets pointed out by GJ15 are reduced with the STE ephemeris.

2014) was used to calculate the positions of Jupiter for the 3 models. We see that the systematic JPL ephemeris offsets pointed out by GJ15 are reduced with our ephemeris, as expected.

In Table 3 we present the mean offsets and the respective standard deviation of the GJ15 positions relative to the same 3 ephemeris as above. We can see that the mean offsets as well as most of their standard deviations are greatly reduced with the STE ephemeris. Of course, by construction, we should expect smaller offsets in the comparison of GJ15 positions with STE. However, it is not obvious that these offsets should be that smaller in comparison with the other ephemeris offsets. Notice that the GJ15 positions come from observations made with very distinct instruments at distant sites located at both Earth hemispheres (good parallax angle coverage), making this set of positions not remarkably distinct than any set of positions that were used in the construction of the other two ephemeris. Thus, these ephemeris offsets comparisons suggest that the accuracy of the STE ephemeris is at least slightly better than that of the other ephemeris, at least for the time span of our satellites' observations. This supports the utility of the STE ephemeris for the next few years, making it one of the best choices to use in stellar occultation predictions in the short future for these satellites.

### 2.2 Phoebe's ephemeris

For the specific case of Phoebe, the ninth satellite of Saturn, we have updated the ephemeris published in Desmars et al. (2013). The new ephemeris (PH15) used the same dynamical model, including the perturbations of the Sun and the eight planets, the eight major satellites of Saturn and the $J_2$ parameter. The observations used to fit the model are identical to Desmars et al. (2013) (including 223 Cassini observations) with additional observations from GJ15, Peng et al. (2015), observations from Minor Planet Circulars between 2012 and 2014 (available on the Natural Satellite Data Center (Arlot & Emelyanov 2009), and observations from Flagstaff (U.S.N.O 2015) between 2012 and 2014. It represents a total number of 5886 observations from 1898 to 2014. In contrast, in Desmars et al. (2013) was used 3367 observations from 1898 to 2012. This represents an increase of almost 75% in the number of observations, mainly with recent observations which is required for our purpose.

---

[1] Last update: February 19, 2012

**Table 2.** Initial osculating elements for Jupiter irregular satellites at JD 2451545.0 with respect to the center of Jupiter.

| Satellite | N | Time-span | a (km) | e | I° | Ω° | ω° | υ° |
|---|---|---|---|---|---|---|---|---|
| Himalia | 1234 | 1995-2014 | 11372100 ± 500 | 0.166 ± 0.002 | 45.14 ± 0.15 | 39.77 ± 0.19 | 351.48 ± 0.46 | 97.35 ± 0.48 |
| Elara | 636 | 1996-2014 | 11741170 ± 690 | 0.222 ± 0.002 | 28.64 ± 0.18 | 68.42 ± 0.43 | 179.82 ± 0.56 | 339.08 ± 0.82 |
| Lysithea | 234 | 1996-2010 | 11739900 ± 1300 | 0.136 ± 0.004 | 51.12 ± 0.27 | 5.53 ± 0.52 | 53.0 ± 1.5 | 318.9 ± 2.0 |
| Leda | 98 | 1996-2009 | 11140300 ± 4300 | 0.173 ± 0.007 | 16.15 ± 0.75 | 272.6 ± 1.7 | 212.2 ± 3.6 | 218.8 ± 3.2 |
| Pasiphae | 609 | 1996-2013 | 23425000 ± 5000 | 0.379 ± 0.001 | 152.44 ± 0.10 | 284.59 ± 0.21 | 135.96 ± 0.19 | 236.97 ± 0.16 |
| Sinope | 221 | 1996-2009 | 22968800 ± 5200 | 0.316 ± 0.002 | 157.76 ± 0.12 | 256.62 ± 0.55 | 298.38 ± 0.55 | 167.57 ± 0.19 |
| Carme | 331 | 1996-2013 | 24202924 ± 4800 | 0.242 ± 0.001 | 147.13 ± 0.10 | 154.01 ± 0.25 | 47.90 ± 0.29 | 234.41 ± 0.19 |
| Ananke | 250 | 1996-2010 | 21683800 ± 7200 | 0.380 ± 0.002 | 172.29 ± 0.20 | 56.9 ± 1.2 | 123.3 ± 1.2 | 231.24 ± 0.21 |

**Notes:** N: number of observations used; a: semimajor axis; e : eccentricity; I: inclination relative to the equatorial reference plane J2000; Ω: longitude of the ascending node; ω: argument of periapsis; υ: true anomaly.

**Table 3.** Mean offsets and standard deviation of the GJ15 positions relative to the STE, Jacobson et al. (2012) and Emelyanov (2005) ephemeris .

| | STE | | JPL | | Eme2008 | |
|---|---|---|---|---|---|---|
| Satellite | $\Delta\alpha\cos\delta$ (mas) | $\Delta\delta$ (mas) | $\Delta\alpha\cos\delta$ (mas) | $\Delta\delta$ (mas) | $\Delta\alpha\cos\delta$ (mas) | $\Delta\delta$ (mas) |
| Himalia | −15 ± 66 | −7 ± 54 | −19 ± 80 | −11 ± 52 | −18 ± 72 | −13 ± 53 |
| Elara | 3 ± 92 | −12 ± 57 | 20 ± 92 | −50 ± 69 | 23 ± 94 | −83 ± 81 |
| Lysithea | 15 ± 79 | −21 ± 68 | 40 ± 92 | −43 ± 77 | 117 ± 193 | −76 ± 185 |
| Leda | −9 ± 67 | −8 ± 77 | 60 ± 117 | −13 ± 95 | 166 ± 162 | 92 ± 95 |
| Pasiphae | 4 ± 89 | −16 ± 57 | −17 ± 130 | −82 ± 85 | −10 ± 102 | −54 ± 74 |
| Sinope | 9 ± 79 | −4 ± 47 | 10 ± 228 | −35 ± 76 | 11 ± 227 | −52 ± 63 |
| Carme | 14 ± 73 | −1 ± 51 | −3 ± 114 | −80 ± 102 | −6 ± 108 | −45 ± 80 |
| Ananke | −10 ± 90 | 3 ± 73 | 60 ± 127 | −108 ± 99 | 101 ± 180 | −107 ± 120 |

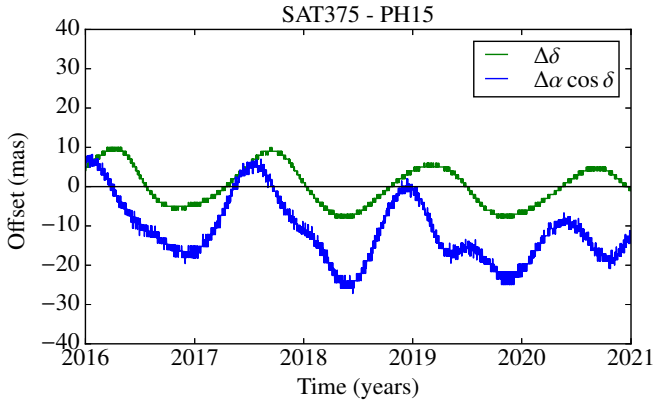

**Figure 2.** Comparison between the PH15 and sat375 JPL ephemeris for the satellite Phoebe.

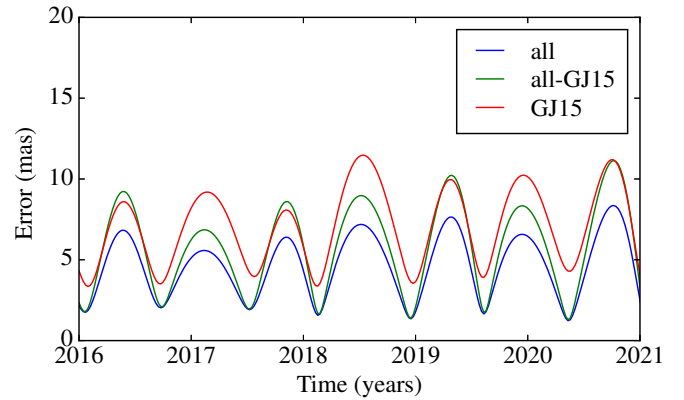

**Figure 3.** Comparison of the precision in on-sky Phoebe-Saturn angular separation for the PH15 ephemeris where three different sets of positions were used to compute the ephemeris: all observations; only positions of GJ15; and all the positions without GJ15.

In Fig. 2 we compare our ephemeris (PH15) with the sat375 JPL[2] ephemeris. The difference between them is smaller than 30 mas (< 10 mas in Declination). This difference is smaller than the apparent diameter of Phoebe (see Table 1)

We computed the precision of the PH15 ephemeris considering three sets of observations: all the positions available; only the positions of GJ15; and all the positions without GJ15. The precision is computed by propagation of the covariance matrix of the orbit determination process and by linear transformations giv-ing the covariance matrix in spherical coordinates (right ascension and declination) at a specific date (for more details, see Desmars et al. 2013). This last matrix then provides the standard deviation in right ascension $\sigma_\alpha$ and declination $\sigma_\delta$ at the required date, with $\sigma_s = \sqrt{\sigma_\alpha^2 \cos^2 \delta + \sigma_\delta^2}$ being the total error in the celestial sphere.

In Fig. 3 we show the comparison between them for the time span 2016-2021 in on-sky Phoebe-Saturn angular separation. It is possible to see that even considering only the positions of GJ15, the estimated error of the ephemeris is smaller than 12 mas. The computed precision does not take into account the precision in the position of Saturn.

**Table 4.** Number of stellar occultations for each satellite from January, 2016 up to December, 2020.

| Satellite | 2016 | 2017 | 2018 | 2019 | 2020 | Total |
|-----------|------|------|------|------|------|-------|
| Ananke    | 12   | 16   | 49   | 359  | 187  | 623   |
| Carme     | 20   | 14   | 30   | 369  | 220  | 653   |
| Elara     | 14   | 16   | 33   | 305  | 193  | 561   |
| Himalia   | 15   | 12   | 54   | 257  | 230  | 568   |
| Leda      | 8    | 24   | 38   | 362  | 208  | 640   |
| Lysithea  | 16   | 11   | 35   | 330  | 212  | 604   |
| Pasiphae  | 20   | 19   | 44   | 362  | 206  | 651   |
| Sinope    | 15   | 21   | 34   | 356  | 256  | 682   |
| Phoebe    | 32   | 98   | 238  | 79   | 13   | 460   |

## 3 PREDICTION OF STELLAR OCCULTATIONS

### 3.1 Candidate events

The prediction of the occultations was made by crossing the stellar coordinates and proper motions of the UCAC4 catalogue (Zacharias et al. 2013) with the ephemeris presented in Sec. 2. The search for stellar candidates follows the same procedure as presented by Assafin et al. (2010, 2012) and Camargo et al. (2014).

We predicted occultations for the 8 major irregular satellites of Jupiter, Ananke, Carme, Elara, Himalia, Leda, Lysithea, Pasiphae and Sinope, and for Phoebe of Saturn.

A total of 5442 events were identified between January 2016 and December 2020. In Table 4 we present the number of stellar occultations predicted by year for each satellite. It is possible to see an increase in the number of events found for Phoebe in 2018 and for the satellites of Jupiter in 2019-2020. This is because at that periods these satellites will cross the apparent galactic plane. We call attention that about 10% of the events will involve stars brighter than magnitude R=14 (and almost 25% brighter than R=15), which helps the attempt of amateur observers.

Table 5 shows a sample of the catalogue of predicted occultations and their parameters, which are necessary to produce occultation maps. Since these objects are very small, the duration of each event is a few seconds. All the occultation tables and maps will be publicly available at the CDS (Centre de Données astronomiques de Strasbourg). In Fig. 4 we show an example of an occultation map. This is an occultation by Elara that will happen in February 21, 2017. This event can be observed from Australia and it is one of the best opportunities for this object due to the slow velocity of the event and it involves a bright star (R*=12.4).

The first preliminary catalogue version of the ESA astrometry satellite GAIA (de Bruijne 2012) is expected to be released up to the end of 2016 (The catalogue with five-parameter astrometric solutions is up to the end of 2017). The precise star positions to be derived by GAIA will provide better predictions with the main source of error being the ephemeris. Astrometric reduction of observations published in GJ15 will be revised with the GAIA catalogue and the predictions will be improved. In that context, in the GAIA era, the occultations predicted will be updated.

### 3.2 Robustness of predictions

Since 2009 many successful observations of stellar occultations by TNOs have been reported in the literature (Elliot et al. 2010; Sicardy et al. 2011; Ortiz et al. 2012; Braga-Ribas et al. 2013), the main disadvantages in their prediction being large heliocentric distances and ephemeris error, facts somewhat compensated

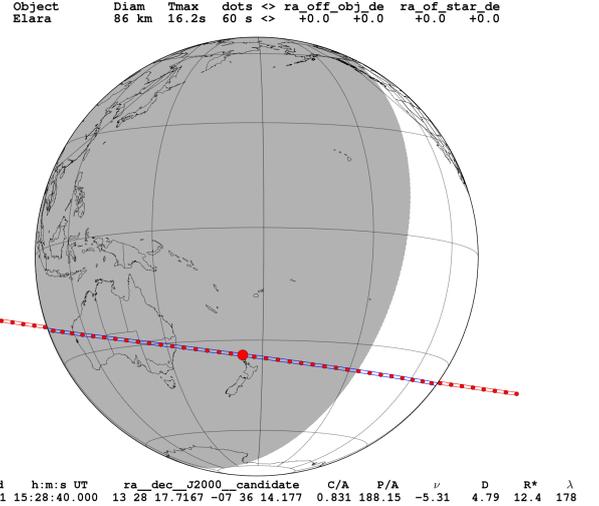

```
Object        Diam    Tmax     dots <>  ra_off_obj_de   ra_of_star_de
Elara         86 km   16.2s    60 s <>    +0.0   +0.0     +0.0   +0.0
```

```
year-m-d      h:m:s UT    ra__dec__J2000_candidate     C/A    P/A      ν       D     R*     λ
2017-02-21  15:28:40.000  13 28 17.7167 -07 36 14.177  0.831 188.15  -5.31   4.79  12.4   178
```

**Figure 4.** Occultation map for Elara. The central red dot shows the geocentric closest approach of the shadow. The small ones show the center of the shadow separated by 60s. The lines show the path of the shadow over the Earth. The shadow moves from right to left. **Labels:** Diam: Diameter of the object; Tmax: Maximum duration of the event for an observation in the center of the shadow; C/A: apparent geocentric distance between the satellite and the star (a.k.a. the apparent distance in the plane of the sky between the shadow and the center of the Earth) at the moment of the geocentric closest approach, in arcseconds; P/A: the satellite position angle with respect to the occulted star at C/A, in degrees; $\nu$: relative velocity of event in km/s; $D$: Geocentric distance to the occulting object in AU; $R^*$: normalized UCAC4 magnitude in the R-band to a common shadow velocity of 20 km s$^{-1}$; $\lambda$: east longitude of subplanet point in degrees, positive towards east, at the central instant of the geocentric closest approach (see the notes of Table 5).

for the larger diameters involved. In contrast to TNOs, the irregular satellites have much better ephemeris because the orbits of their host planets are better known, their observational time span is much wider and covers many orbital periods. Moreover, the irregular satellites are much closer to Earth which implies in a much smaller shadow path error in kilometers. These advantages may be somewhat balanced by the smaller sizes estimated for the irregular satellites. Thus, in comparison, the chances for a successful observation of an stellar occultation by an irregular satellite should be considered at least also as good as those by TNOs.

Observing a stellar occultation demands a great effort. And, in our case, the shadow covers a very restricted area on Earth because of the size of the irregular satellites. Since no stellar occultation by an irregular satellite was observed up to date, and since we want to be sure that we can start observational campaigns with reasonable chances of success, we tested the robustness of an occultation prediction for a large target.

The test design consisted in observing the object and star to be occulted near the date of the event predicted when the two objects were present in the same field of view (FOV), close to each other. Thus, the relative positions between the two objects had minimal influence of the errors of the reference catalogue of stars used and possible field distortions (Peng et al. 2008, and references therein). The relative positions of the star and satellite were used to check the original prediction. Notice that in the test we did not attempt to observe any actual occultation. The test could be performed at any site, regardless of the Earth location where the occultation would in fact be visible.

**Table 5.** A sample of stellar occultation predictions for Pasiphae

| d m Year   h m  s | RA  (ICRS)  Dec | C/A | P/A | $\nu$ | D | $R^*$ | $\lambda$ | LST | $\mu_{\alpha*}$ | $\mu_\delta$ |
|---|---|---|---|---|---|---|---|---|---|---|
| 09 04 2016 03:58:19. | 11 14 36.7707 +07 39 20.7610 | 1.003 | 17.9 | -12.88 | 4.54 | 14.9 | 271. | 22:03 | 12. | -33. |
| 13 06 2016 00:16:12. | 11 12 48.5020 +07 06 43.3520 | 0.661 | 30.0 | +14.32 | 5.50 | 13.9 | 262. | 17:45 | -1. | 1. |
| 27 06 2016 13:56:09. | 11 18 03.4160 +06 23 45.1940 | 1.707 | 28.0 | +20.29 | 5.74 | 11.7 | 44. | 16:53 | 4. | -10. |
| 18 07 2016 15:07:24. | 11 28 15.5076 +05 05 31.8060 | 0.942 | 26.7 | +27.80 | 6.05 | 14.0 | 8. | 15:40 | 4. | 4. |
| 22 07 2016 16:15:07. | 11 30 30.4310 +04 48 43.4340 | 0.644 | 206.5 | +29.04 | 6.11 | 14.6 | 348. | 15:27 | 23. | -24. |
| 24 07 2016 01:37:34. | 11 31 17.8471 +04 42 49.0540 | 0.029 | 206.6 | +29.46 | 6.12 | 15.1 | 206. | 15:22 | 2. | -8. |
| 24 07 2016 17:37:18. | 11 31 40.7472 +04 39 57.5060 | 0.840 | 26.5 | +29.66 | 6.13 | 14.9 | 326. | 15:20 | -11. | -1. |

**Notes**. Entries included: day of the year and UTC central instant of the prediction; right ascension and declination of the occulted star - at the central instant of the occultation (corrected by proper motions); C/A: apparent geocentric distance between the satellite and the star (a.k.a. the distance between the shadow and the center of the Earth) at the moment of the geocentric closest approach, in arcseconds; P/A: the satellite position angle with respect to the occulted star at C/A, in degrees (zero at north of the star, increasing clockwise); $\nu$: relative velocity of event in km s$^{-1}$: positive = prograde, negative = retrograde; D: Geocentric distance to the occulting object in AU; $R^*$: normalized UCAC4 magnitude in the R-band to a common shadow velocity of 20 km s$^{-1}$ by the relationship $R^* = R_{UCAC4} + 2.5 \times \log 10 \left( \frac{velocity}{20km s^{-1}} \right)$, the value 20 km s$^{-1}$ is typical of events around the opposition; $\lambda$: east longitude of subplanet point in degrees, positive towards east, at the instant of the geocentric closest approach; LST: UT + $\lambda$: local solar time at subplanet point, hh:mm; $\mu_{\alpha*}$ and $\mu_\delta$: proper motions in right ascension and declination, respectively (mas/year). For more detailed information about the definition and use of these stellar occultation geometric elements see Assafin et al. (2010).

We tested the occultation by Himalia predicted to occur on March 3, 2015. The shadow would cross the northern part of South America. For the event, four situations were considered:

(i) Our nominal, published prediction with the STE ephemeris (see Sec. 2), and the nominal UCAC4 position of the star;

(ii) Prediction with the JPL ephemeris and the nominal UCAC4 position of the star;

(iii) Prediction from star and satellite offsets calculated from observations made a few days before the occultation when the objects were very separated (different FOVs);

(iv) Same as (iii) but with the star and the satellite close in the same FOV.

Table 6 shows the differences between the predictions in the four situations. For situation (iii) we observed the objects on February 22 with the Zeiss telescope (diameter = 0.6m; FOV = 12.'6; pixel scale = 0.''37/pixel) at the Observatório do Pico dos Dias, Brazil (OPD, IAU code 874, 45°34′57″W, 22°32′04″S, 1864m). On that day, Himalia and the star were observed in separate FOVs as they were still far apart. On the night of the event, March 3, the objects were observed with Perkin-Elmer telescope (diameter = 1.6m; FOV = 5.'8; pixel scale = 0.''17/pixel) at OPD just over an hour after the time scheduled for the event. Satellite and star were separated by about 16 arcsec, so very close to each other (situation (iv)). From the calculated offsets, the center of the shadow was obtained. Notice that the shadow path was not predicted to cross the OPD (which was located at almost 2000 km south from the shadow path). This was not necessary for testing the prediction.

The critical parameter in the comparisons is the C/A, which here is related to latitudes. The apparent radius of Himalia is about 20 mas (see Table 1). In the context of the test, for a 0 mas offset in C/A we would have 100% probability of observing the occultation, and 0% in the case of a C/A offset equal to or larger than 20 mas, the radius of Himalia. From Table 6, we have nearly 0% probability of success in situation (iii), for which the offset in C/A was -20 mas, but when the relative astrometry was poor, 10 days prior to the event. Once at the day of the event in situation (iv), the C/A offset dropped to -9 mas only, corresponding to a 55% probability of success. Comparison with the prediction using the JPL ephemeris (situation (ii)) gives a +11 mas C/A offset, or a compatibility of

**Table 6.** Comparison between the predictions of the Himalia occultation at March 03, 2015.

| Differences with respect to the STE prediction | | | |
|---|---|---|---|
| Method | Instant of C/A | C/A | Sit. |
| STE | 00:39:51 UTC | 0.''703 | (i) |
| JPL | -26s | +11mas (36km) | (ii) |
| Feb. 22 Obs. | -14s | -20mas (65km) | (iii) |
| Mar. 03 Obs. | -36s | -09mas (29km) | (iv) |

C/A: geocentric closest approach; Sit: Situation test considered.

45% between the ephemerides. All this suggests that there was a good probability of observing the event. The largest differences between the shadows of the four situations were 36s in time along the shadow path and 101km (31 mas) in the direction perpendicular to the shadows, suggesting that observers should be spread in narrow latitude ranges 100 km wide.

## 4 DISCUSSION

We performed new numerical integrations for improving the orbits of some of the larger irregular satellites. Consequently, with our ephemeris, we predicted stellar occultations aiming to access fundamental parameters like size, shape, albedo, ultimately aiming to track the formation origin of these bodies.

For the irregular satellites of Jupiter (Ananke, Carme, Elara, Himalia, Leda, Lysithea, Pasiphae and Sinope), we produced ephemeris using only the observations of GJ15. These new ephemerides are denominated Special-Tailored Ephemeris.

We also updated the ephemeris of Phoebe (Desmars et al. 2013) using the observations of GJ15, Peng et al. (2015), observations from MPC and from Flagstaff. A total of 5886 observations between 1989 and 2014 were used in the process. This represents and increase of about 75% in the number used to generate the ephemeris of Phoebe in Desmars et al. (2013).

As it was shown for Phoebe, when we use only the positions of GJ15, the ephemeris presents a precision in the order of those where all the positions were used. Moreover, the case of Phoebe is particular because we have many observations in a large time spam (1898 to 2014) including observations from Cassini. For the Jovian satellites with less observations, the precision we have by using only GJ15 observations may be quite equivalent or even better than the precision of other ephemeris for the short time span explored in this work.

We predict stellar occultations for the period of 2016-2020 for eight irregular satellites of Jupiter: Ananke, Carme, Elara, Himalia, Leda, Lysithea, Pasiphae, and Sinope; and one satellite of Saturn: Phoebe. The procedure used was the same as that for the prediction of stellar occultations by Pluto and its satellites in Assafin et al. (2010) and by Centaurs and TNOs in Assafin et al. (2012) and Camargo et al. (2014).The candidate stars were searched in the UCAC4 catalogue. The occasional passage of Jupiter by the galactic plane in 2019-2020 and Saturn in 2018 creates the best opportunity of observing stellar occultations in the near future due to the great density of stars in the region. Indeed, a total of 5442 events are foreseen. The next time that Jupiter will cross the central side of the galactic plane will be in 2031 and Saturn in 2046-2047.

In a broader, general sense, the probability of successfully observing an occultation is roughly the ratio of the satellite's radius by the budget error (2 sigma for a 95% confidence level) of ephemeris and star position. Thus, UCAC4 errors ranging between 20 mas - 50 mas (1 sigma) combined with a mean error (1 sigma) in the JPL ephemeris of 30 mas for Himalia and 150 mas for Leda published in Table 2 of Jacobson et al. (2012) would give 28%-17% probability of observing such an event by Himalia and ≈ 2% for Leda, the smallest irregular satellite in the sample. Observations a few days before the date of occultation predicted may improve the combined errors to 40-80 mas, depending on the magnitude of the objects. This probability is estimated for a single observing site, and we expect to reach higher probability with multi-sites.

The test made with an occultation expected to happen in March 03, 2015 for Himalia showed that this event would probably have been observed successfully in case there were observers available in the shadow area. The results show satisfying small offsets with respect to the local of the prediction.

GJ15 also observed Sycorax (satellite of Uranus) and Nereid (satellite of Neptune). There were few observations of Sycorax distributed in 9 nights over two years which did not cover one orbital period. For Nereid, the observations covered many orbital periods, but due to Nereid's large orbital eccentricity there are no observations near the pericenter.

Uranus and Neptune are crossing a very low dense region of stars. This results in almost no stellar occultation by these objects up to 2020. In fact, using JPL ephemeris, we identified only 2 events for each satellite in this period, but due to the bad conditions of the events (shadow far from observatories; faint stars) we chose not to publish any events here. For these reasons we did not attempt to generate new orbits for these satellites here.

Continuous observations of the satellites are recommended and fitting of our dynamical model to those observations are expected to reduce the respective STE ephemeris errors. The first version of the GAIA catalogue is to be released up to the end of 2016 and will improve the position error of the stars to the 1-5 mas level. Re-reduction of older positions, and reduction of new positions of irregular satellites with GAIA will improve new orbit determinations. It will also allow for the discovery of occultations by more stars not present in the UCAC4 catalogue. The release of the GAIA catalogue should have a positive impact on both the astrometric precision of occulted stars and the reduction of new astrometric positions of the satellites. As a result, prediction of stellar occulta-tions by irregular satellites shall increase in number as well as in precision and success.

## ACKNOWLEDGEMENTS

ARG-J thanks the financial support of CAPES. MA thanks the CNPq (Grants 473002/2013-2 and 308721/2011-0) and FAPERJ (Grant E-26/111.488/2013). RV-M thanks grants: CNPq-306885/2013, Capes/Cofecub-2506/2015, Faperj: PAPDRJ-45/2013 and Cient.Est-05/2015. JIBC acknowledges CNPq for a PQ2 fellowship (process number 308489/2013-6). BEM thanks the financial support of CAPES. FB-R acknowledges PAPDRJ-FAPERJ/CAPES E-43/2013 number 144997, E-26/101.375/2014. The numerical model of the satellites of Jupiter was developped during a post-doctoral contract funded by the Chinese Academy of Sciences (CAS) and supported by the National Scientific Fund of China (NSFC). The authors thank Dr. Nikolay Emelyanov for his comments to improve the paper.